# Element Abundances and Source Plasma Temperatures of Solar Energetic Particles

**Donald V Reames**

IPST, Univ. of Maryland, College Park, MD 20742-2431 USA
dvreames@umd.edu

**Abstract**. Thirty years ago Breneman and Stone [1] observed that the enhancement or suppression of element abundances in large solar energetic-particle (SEP) events varies as a power of the mass-to-charge ratio, *A/Q*, of the elements. Since *Q* during acceleration or transport may depend upon the source plasma temperature *T*, the pattern of element enhancements can provide a best-fit measure of *T*. The small SEP events we call $^3$He-rich or "impulsive" show average enhancements, relative to coronal abundances, rising as the 3.6 power of *A/Q* to a factor of ~1000 for (76≤*Z*≤82)/O and temperature in the range 2-4 MK. This acceleration is believed to occur in islands of magnetic reconnection on open field lines in solar flares and jets. It has been recently found that the large shock-accelerated "gradual" SEP events have a broad range of source plasma temperatures; 69% have coronal temperatures of *T* < 1.6 MK, while 24% have *T* ~ 3 MK, the latter suggesting a seed population containing residual impulsive suprathermal ions. Most of the large event-to-event abundance variations and their time variation are largely explained by variations in *T* magnified by *A/Q*-dependent fractionation during transport. However, the non-thermal variance of impulsive SEP events (~30%) exceeds that of the ~3 MK gradual events (~10%) so that several small impulsive events must be averaged together with the ambient plasma to form the seed population for shock acceleration in these events.

## 1. Introduction

The relative abundances of the chemical elements vary greatly among the energetic particles from the Sun and the study of those abundances has revealed many underlying physical processes involved in their acceleration and transport. Most of these processes can only be studied using direct samples of the particles themselves, since photon production by accelerated ions is extremely limited and is poorly observed.

Logically the primary distinction was between SEP events we call "gradual" and "impulsive" [2, 3, 4, 5, 6, 7]. Historically, gradual events, being the most intense, were observed first. They are now known to be accelerated at shock waves driven out from the Sun by coronal mass ejections (CMEs). Most of the acceleration begins at 2 to 3 solar radii and samples ambient coronal plasma. It was recognized by Meyer [8] that the abundances, averaged over large SEP events and divided by the corresponding photospheric abundances shows a characteristic two-level pattern when plotted *vs.* the first ionization potential (FIP) of the element, as shown in the left panel of Figure 1. This occurs because elements with low FIP are ionized in the photosphere while those with high FIP are neutral so the former are much more easily transported up into the corona. Gradual SEP events, on average, measure coronal abundances. However a second important aspect of gradual event abundances, also shown in Figure 1 was that in individual events, abundances relative to the mean, varied as a power law in the charge-to-mass ratio ($Q/M$ or $Q/A$) of the ions.

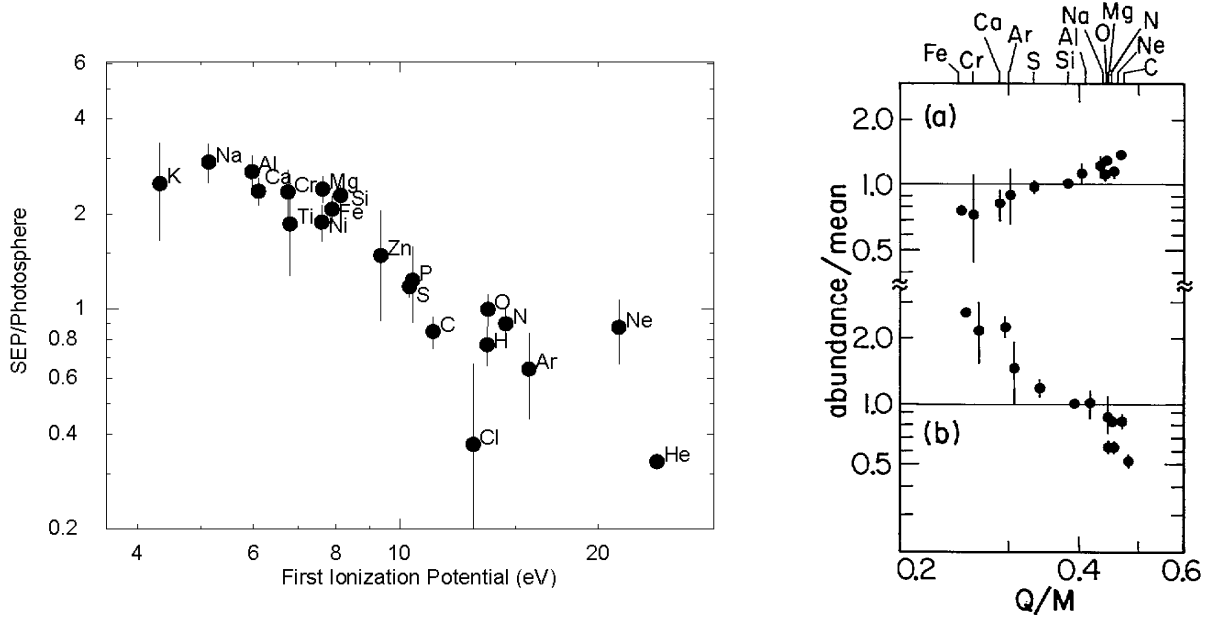

**Figure 1.** The left panel shows the average SEP abundances [9] relative to corresponding photospheric abundances [10] *vs.* FIP. The right panel shows $Q/M$ dependence for two large events observed by Breneman and Stone [1].

It might be said that impulsive events were first distinguished by Wild, Smerd, and Weiss [11] as radio type III bursts produced by 10–100 keV electrons streaming out from the Sun. However the associated ions, found much later, turned out to have strange abundances, beginning with greatly enhanced $^3He/^4He$ ratios [12], exceeding that of the corona or solar wind by factors up to $10^4$ [13]. These $^3$He-rich events were eventually associated with the electron-rich type III bursts [13] and the heavier ions were found to have $A/Q$-dependent enhancements of a factor of ~10 at Fe [14]. These abundances led to the first estimate of source plasma temperature $T$, since He, C, N, and O appeared to be fully ionized, Ne, Mg, and Si had two orbital electrons, and Fe had a higher $A/Q$, consistent with $T$ = 3–5 MK. Impulsive SEP events have been associated with magnetic reconnection on open field lines as occur in solar jets [15]. Particle-in-cell simulations support strong power-laws in $A/Q$ [16] and the $^3$He enhancements are believed to result from preferential absorption by $^3$He of electromagnetic ion-cyclotron waves generated by the streaming electrons [17].

Our ability to cleanly distinguish impulsive and gradual SEP events using element abundances alone is prevented by the fact that shock waves frequently encounter residual suprathermal ions left over from impulsive SEP events among the seed population that becomes accelerated. Enhanced abundances of $^3$He were found in large shock events [18] and are seen before and after shock acceleration at *in situ* observations [19]. Impulsive suprathermal ions can alter the energy dependence of abundance ratios such as Fe/O [20].

**2. Element abundances and source plasma temperatures in impulsive SEP events**
Recent measurements of ions in impulsive SEP events using two different instrument techniques [21] have extended the abundance measurements to elements throughout the periodic table (although not with single-element resolution). Figure 2 shows the abundance enhancement, relative to the SEP coronal abundances, averaged over 111 impulsive SEP events as a function of $A/Q$ of the ions.

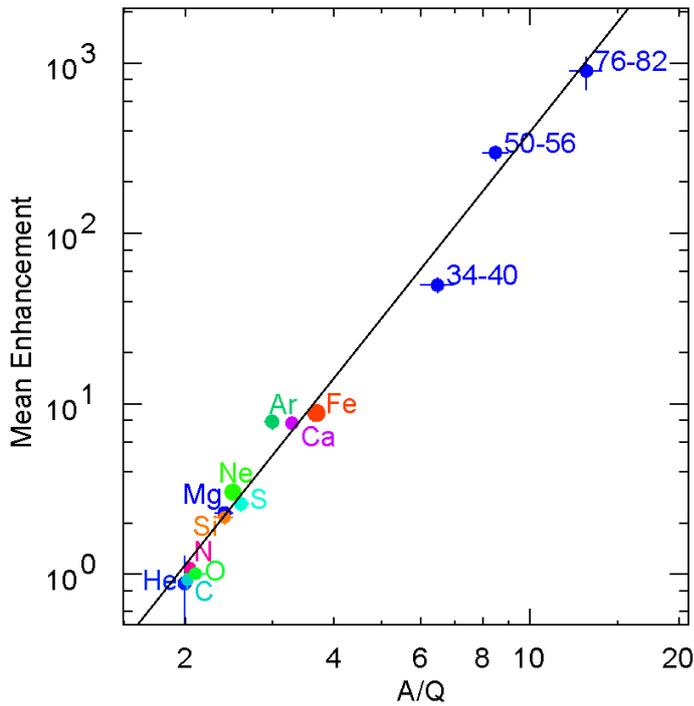

**Figure 2**. The mean element enhancement in impulsive SEP events (relative to coronal abundances) is shown *vs.* *A/Q* at 2.5–3.2 MK where values are taken from the shaded band in Figure 3. The power of the fitted line shown is 3.64±0.15 [22].

Figure 3 shows *A/Q vs.* temperature *T* using theoretical values of *Q vs. T* [23] for typical elements.

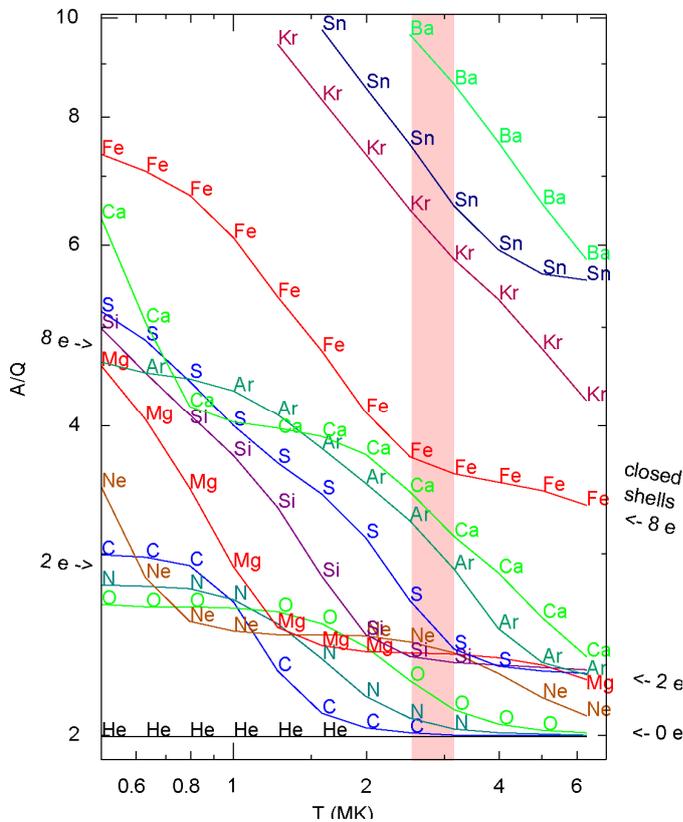

**Figure 3**. *A/Q* is plotted *vs.* the theoretical equilibrium temperature for elements that are named along each curve [23]. Points are spaced every 0.1 unit of log *T* from 5.7 to 6.8. Bands produced by closed electron shells with 0, 2, and 8 electrons are indicated in the margins, He having no electrons. Elements tend to move from one of these groups to another as the temperature changes. The red strip selects values used for impulsive SEP events in Figure 2 [23, 26].

Early observations [24] were ascribed to the range of 3–5 MK where He, C, N, and O are fully ionized and Ne, Mg, and Si have 2 orbital electrons. However, the lower range better fits the heavy elements, explains why Ne is enhanced more than Mg or Si, and explains the existence of "C-poor" and "He-poor" events where O has been enhanced so that He/O and C/O decrease [22]. The strong power-law dependence on *A/Q* agrees with theoretical particle-in-cell simulations of collapsing islands of magnetic reconnection [16].

It is also possible to estimate *T* for individual impulsive SEP events by fitting the observed enhancements *vs. A/Q(T)* for each of several temperatures of interest [25]. The fit with the lowest value of $\chi^2$ is selected. Nearly all of the impulsive events have minimum values of $\chi^2$ at either 2.5 or 3.2 MK. Most (69%) of the events associate with CMEs, most *slow* and *narrow*, suggesting solar jets, but, otherwise, the SEP parameters are not strongly correlated with CME or flare parameters [22, 25].

## 3. Element abundances and source plasma temperatures in gradual SEP events

Breneman and Stone [1] showed that abundance enhancements in gradual event also behave as a power law in *A/Q*. It can be shown from diffusion theory that the scattering of ions during transport can produce an enhancement (or suppression) of the abundance ratios that is a power law in *A/Q* with a power that varies with time *t* during the event as $a/t + b$ [26]. Here the parameter *b* is negative unless previously enhanced impulsive suprathermal ions are accelerated.

Figure 4 shows the analysis of several energy intervals in a large gradual event. Fits at each temperature yield values of $\chi^2$ and the temperature of the minimum $\chi^2$ is selected.

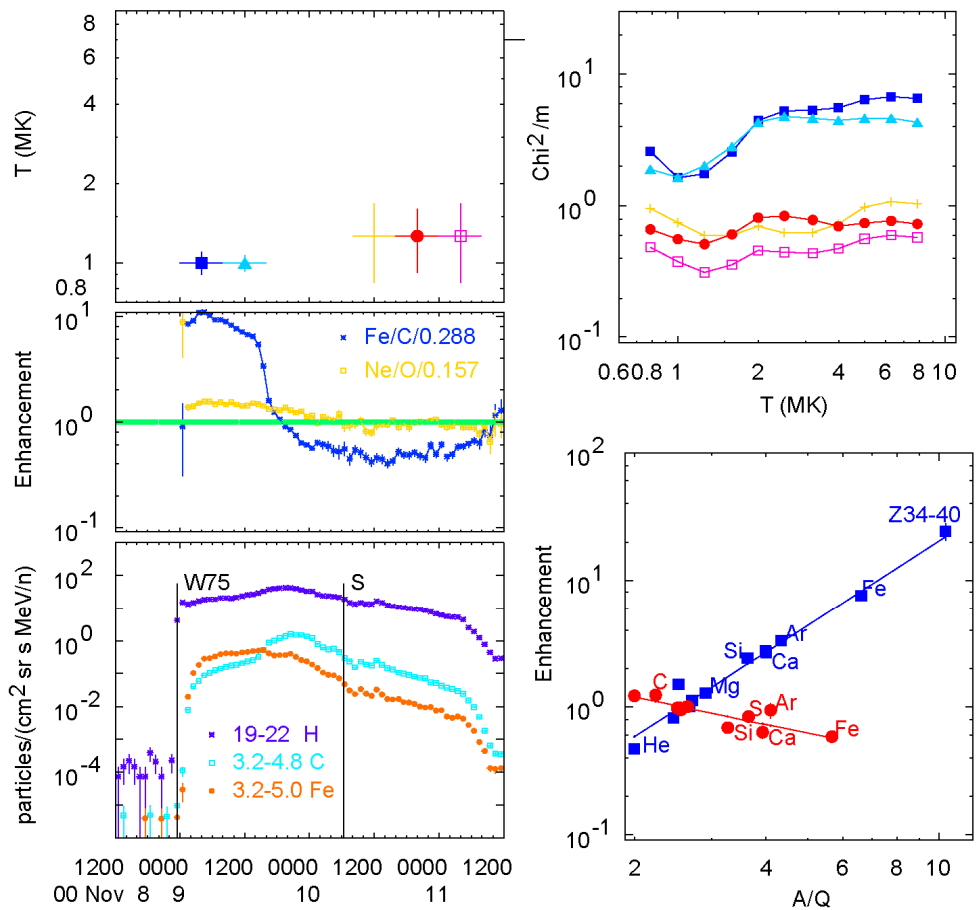

**Figure 4.** Clockwise from the lower left panel are the intensities of H, C, and Fe during the 8 November 2000 SEP event, the enhancements in Fe and Ne during the event, the best-fit temperatures in color-coded 8-hr intervals, values of $\chi^2/m$ *vs. T* for each time interval, and best power-law fits of the observed enhancements *vs. A/Q* at two times [26].

The determination of temperatures is easily understood by returning to Figure 3. Proceeding downward in temperature, one sees the elements O, then N, then C cross from stripped to the two-electron shell. Meanwhile, the elements Ca, Ar, S, Si, then Mg cross from the 2- to the 8-electron shells. Since the observed enhancements mirror *A/Q*, the pattern of enhancement and the grouping of the elements shows the temperature. Of course, we could scale Fe/O to fit *A/Q* at any *T*, but the intermediate elements group poorly and raise $\chi^2$ except at the best-fit temperature.

Figure 5 compares the pattern of enhancements in the first 8-hr interval of the 2000 November 8 event seen in Figure 3. In the figure the observed enhancements of C, N, and O have moved up above He to be near to Ne and Mg. Meanwhile Si and S have moved up near Ca, Fe is considerably enhanced and we even have a measure of $34 \leq Z \leq 40$, represented by Kr.

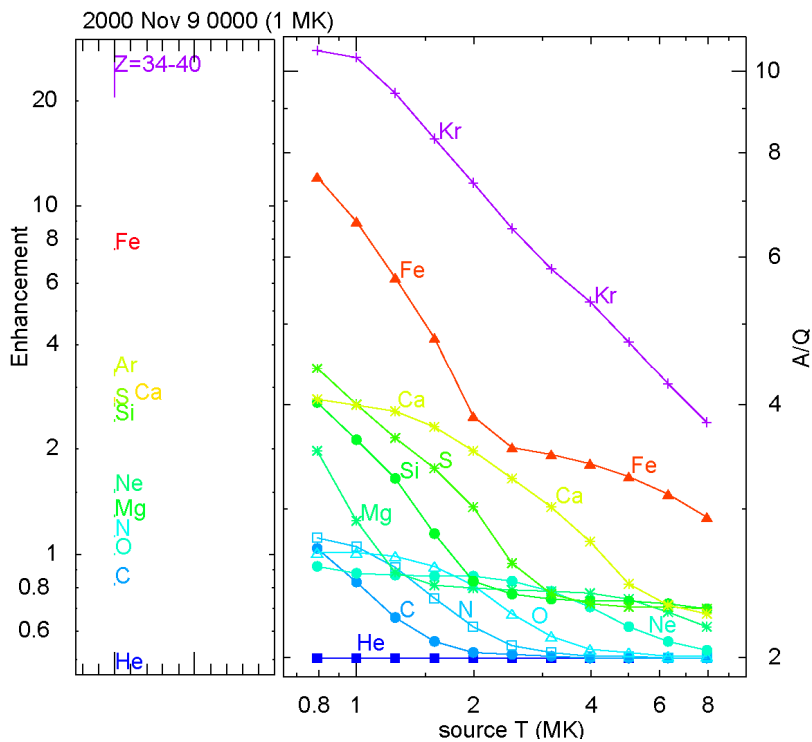

**Figure 5**. The left panel shows enhancements in element abundances during the interval 0000–0800 UT 9 November 2000. The right panel shows A/Q vs. T for various elements, as in Figure 3. The groupings of enhancements match those in A/Q near 1 MK [26].

Out of 45 gradual events that have been studied [26]:

- 11 events (24%) showed source plasma temperatures of 2.5–3.2 MK similar to impulsive SEPs,

- 31 events (69%) had source plasma temperatures of ≤ 1.6 MK,

For about 10% of the latter events, the values of $\chi^2/m$ turn down again at high temperatures. These events are surely not hot plasma but may be a component that has been stripped of electrons.

Generally correlations between source plasma temperatures and properties of the associated CME are rather weak. Figure 6 shows *T vs.* CME speed which has an un-weighted correlation coefficient of -0.49. Also, several events in the sample are ground level events (GLEs) in which GeV protons cause a nuclear cascade in the atmosphere that is measurable in instruments at ground level. These GLEs are identified in Figure 6. However, there seems to be nothing special about them.

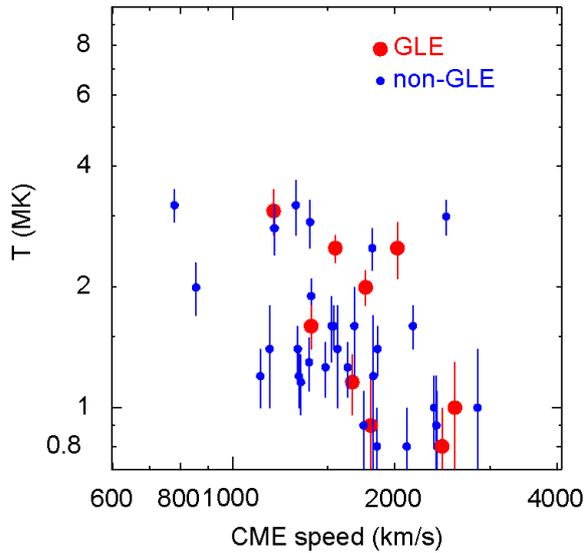

**Figure 6**. Source plasma temperatures of the SEP events are shown as a function of speed of the associated CME. GLEs among the sample, indicated as large red circles, have no special distribution [26].

Temperature differences explain many of the abundance variations between gradual SEP events that were previously unexplained. However, interesting new differences between impulsive and gradual events with the same source temperatures have also emerged as seen in Figure 7.

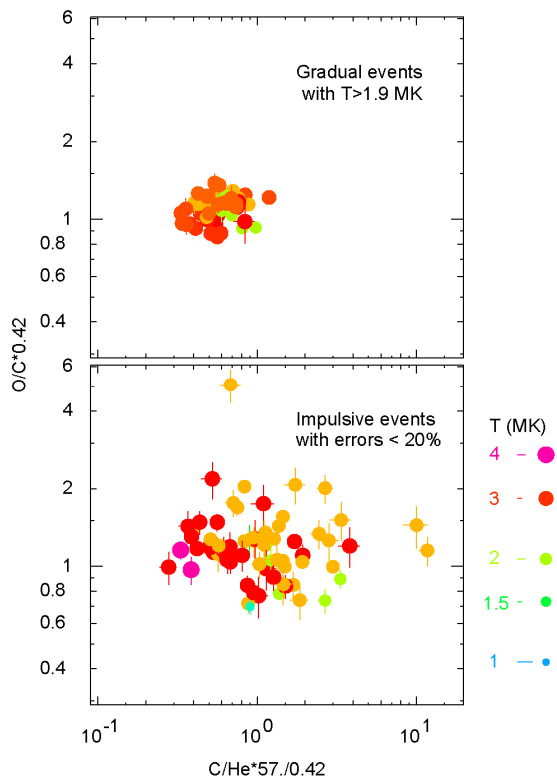

**Figure 7.** The figure compares normalized O/C *vs.* C/He for gradual SEP events with $T \geq 2$ MK (upper panel) and for well-measured impulsive SEP events (lower panel) at the same scale. The impulsive events show large non-thermal errors that must be reduced by averaging in the gradual events.

At the temperatures of impulsive SEPs, ~3 MK, He, C, and possibly even O, should be almost fully ionized so that no relative enhancements are possible. Thus the scatter of the impulsive events in this figure must come from *non-thermal* variations, possibly abundance variations in the underlying local plasma before acceleration. The comparison in Figure 7 suggests that *no single impulsive event can provide all seed particles for a gradual event*; if it did, the gradual events would have a spread like the impulsive events. However, averaging over many impulsive events would reduce the spread.

Averaging 10 impulsive events with 30% errors would reduce the error in the mean to 10%. Accelerating ambient active region plasma along with impulsive seed particles is also likely and would reduce the errors, but would also reduce Fe/O and Ne/O, possibly more than observed.

Where are the 10 impulsive events that contribute suprathermal seeds to each gradual SEP event? In 1984, Lin *et al.* [27] extended the size distribution of solar X-ray flares to smaller sizes using a large balloon-bourn detector; as the flare energy decreased the number of flares per day increased as a power law. This observation led Parker to suggest nanoflares [28] to heat the corona. Now flares involve magnetic reconnection on closed loops, capturing the energy as heat; we are interested in jets with reconnection on open field lines. Whether or not nanoflares are sufficiently numerous to heat the corona, we suspect that small *nanojets*, erupting frequently in active regions can supply a steady multi-jet seed population of impulsive suprathermal ions that would be available to the shock wave of a gradual event passing through. If the power of the size distribution were 1.72 [27], there would be over 50 times as many events, each contributing 10% of the suprathermal ions accelerated in the impulsive SEP events we see. The nanojets would sample different points in an active region.

Thus we see two situations for the production of a gradual SEP event:

- If an observer is magnetically connected to the *ambient corona* as a big shock goes through, SEPs come from seed ions at the local coronal temperature.

- If an observer is magnetically connected to an *active region* as a big shock goes through, SEP seeds average over the current multi-jet population of impulsive suprathermal ions as well as the sampling some of the ambient corona.

Ko *et al.* [29] found that Fe-rich gradual SEP events were commonly connected to active regions and observations of persistent and long-lived $^3$He-rich sources have been observed [30].

**4. Summary**
Element abundances averaged over many of the large gradual events provide a measure of coronal abundances that differ from photospheric abundances by their well-known dependence on FIP. These SEP coronal abundances serve as a reference for both impulsive and gradual events.

Enhancements of the average abundances in impulsive events, relative to the reference corona, show a strong power-law in *A/Q* to the 3.6 power, determined at a temperature near 3 MK, extending from He to Pb. Such a power-law is obtained theoretically from simulations of islands of magnetic reconnection. These reconnections occur on open magnetic field lines in solar jets and are associated with slow, narrow CME.

We can find a temperature for individual impulsive events by fitting the observed enhancement vs. *A/Q(T)* at several values of *T* and selecting the fit and *T* with the minimum $\chi^2$. Nearly all impulsive events fall at 2.5 or 3.2 MK.

Element abundances in gradual SEP events, relative to the reference corona, show either increasing or decreasing power laws in *A/Q*, as found originally by Breneman and Stone [1] and are expected from diffusive transport theory [26]. Values of *T* can again be found by fitting the observed enhancement vs. *A/Q(T)* at many values of *T* and selecting the fit and *T* with the minimum $\chi^2$. This works because the grouping of *A/Q* values around closed shells is quite clear and varies strongly with *T*. Most (69%) of the gradual events have temperatures < 1.6 MK representing shock acceleration of ambient coronal material, 24% have *T* > 2 MK, probably from shock reacceleration of impulsive suprathermal ions contributed by multiple small nanojets as well as active-region plasma which dilutes the enhancements.